\documentclass[useAMS,usenatbib]{mnras}

\usepackage[utf8]{inputenc}
\usepackage{graphicx}   % Including figure files
\usepackage{amsmath}    % Advanced maths commands
\usepackage{amssymb}    % Extra maths symbols
\usepackage{multicol}   % Multi-column entries in tables
\usepackage{bm}         % Bold maths symbols, including upright Greek
\usepackage{pdflscape}  % Landscape pages
\usepackage{float}      % can't remember
\usepackage{tipa}       % to allow g_{ff} to be written properly
\usepackage{orcidlink}
\usepackage{color, soul}
\soulregister\cite7
\soulregister\ref7
\soulregister\pageref7

\newcommand{\noopsort}[1]{}

\hypersetup{final}
\title[]{Dust cloud lifetimes of Scallop-shell stars}
\author[]{Daley-Yates S.}

\author[Simon Daley-Yates \& Moira M. Jardine \& Luke Bouma]{
    Simon Daley-Yates$^{1, 2}$\thanks{E-mail: sddy1@st-andrews.ac.uk} \orcidlink{0000-0002-0461-3029}, 
    Moira M. Jardine$^{2}$ \orcidlink{0000-0002-1466-5236}, 
    Luke Bouma$^{3}$ \orcidlink{0000-0002-0514-5538} \\   
    $^{1}$School of Mathematics and Statistics, University of St Andrews, North Haugh, St Andrews, Fife, Scotland KY16 YSS, UK \\
    $^{2}$School of Physics and Astronomy, University of St Andrews, North Haugh, St Andrews, Fife, Scotland KY16 YSS, UK,\\
    $^{3}$Observatories of the Carnegie Institution for Science, Pasadena, CA 91101, USA
}

\begin{document}

\date{}

\pagerange{\pageref{firstpage}--\pageref{lastpage}} \pubyear{2023}

\maketitle

\label{firstpage}

\begin{abstract}
We investigate the survival of dust trapped in magnetically confined cool gas clouds (or {\it prominences}) around rapidly rotating M-dwarfs exhibiting the ``scallop-shell'' light-curve morphology. Using a two-dimensional magnetohydrodynamic simulation, we extend previous coronal prominence models to include a passive tracer field to allow for a single injection of collisionally charged dust grains. The tracer evolution reveals how recurrent centrifugal breakouts--the slingshot process--remove dust and gas from the prominence while chromospheric evaporation replenishes gas from below. For our simulated star, which has $R_{\ast} = 0.6 R_{\odot}$, $M_{\ast} = 0.3 M_{\odot}$, and $P_{\ast} = 0.32$ days, the resulting dust content decays exponentially with a minimum half-life of approximately 6 stellar rotations, representing a lower limit set by our assumption of fully coupled dust and gas dynamics. Synthetic velocity-phase diagnostics show a single, phase-locked feature that fades steadily, reproducing the behaviour of dips seen in TESS and K2 light curves. Comparison with observed river plots  suggests a natural classification: (i) persistent, non-decaying features formed by quiescent prominences below co-rotation; (ii) gradually fading features produced by slingshot prominences near co-rotation; and (iii) abrupt disappearances linked to magnetic reconnection and flare-driven ejections. These results demonstrate that dust-bearing prominences--undergoing repeated slingshots--can persist for tens of rotations, linking the observed longevity of the scallop-shell photometric features with the dynamic cycle of prominence slingshot ejections.
\end{abstract}

\begin{keywords}
stars: coronae - stars:  magnetic field - stars: activity
\end{keywords}

\section{Introduction}

% \SDYtext{This is text.}
% \SDYcomment{This is a comment.}

The discovery of ``scallop-shell'' stars (also known as Complex Periodic Variables (CPVs)) has introduced a rich and complex dimension to our understanding of stellar variability among young, rapidly rotating M-dwarfs. These stars, first brought to note through observations from the K2 mission \citep{Rebull2016, Stauffer2017}, exhibit light curves characterized by broad, quasi-sinusoidal modulations superimposed with sharp, near-periodic flux dips. Unlike the variability typically associated with rotational modulation by star spots, these features are narrow and present with remarkable periodic coherence over time spans of weeks to years. Their disappearance can be gradual or, at times often linked to energetic flaring, sudden. The persistent and intricate structure of these photometric signatures has challenged conventional explanations and motivated an extensive body of observational and theoretical research {(see \citealt{Bouma2024, Bouma2025} and references therein for a comprehensive overview).}

Following the discovery of these features, subsequent studies sought to constrain the physical origin of the scallop-shell phenomenon. Early interpretations posited that the dips arise from occultations by gas clouds trapped in large-scale stellar magnetic fields, analogous to the slingshot prominences observed in rapidly rotating stars such as AB Doradus \citep{CollierCameron1989, Jardine2001, Donati2008}. These prominences are known to form along magnetic field lines extending beyond the stellar co-rotation radius, providing stable sites for cool material to gather. However, the absorption depths observed in scallop-shell light curves exceed what would be expected from coronal gas alone. This has led to increasing focus on particulate matter as a contributing factor (see the work of \cite{Sanderson2023} and references therein). High-precision observations with TESS \citep{Zhan2019, Gunther2022}, combined with earlier rotational studies of M dwarfs \citep{Irwin2011, Newton2016, Reiners2012}, and long-term photometric monitoring campaigns \citep{Davenport2015, Stelzer2016} have further refined our understanding of these systems, revealing the long-term persistence of rapid rotation in these stars and showing that the dips are often quasi-stable over tens to hundreds of rotation cycles, yet subject to abrupt disappearance or phase shifts.

\begin{figure*}[t]
	\centering
	\includegraphics[width=0.69\textwidth,trim={0cm, 0cm, 0cm, 0cm},clip]{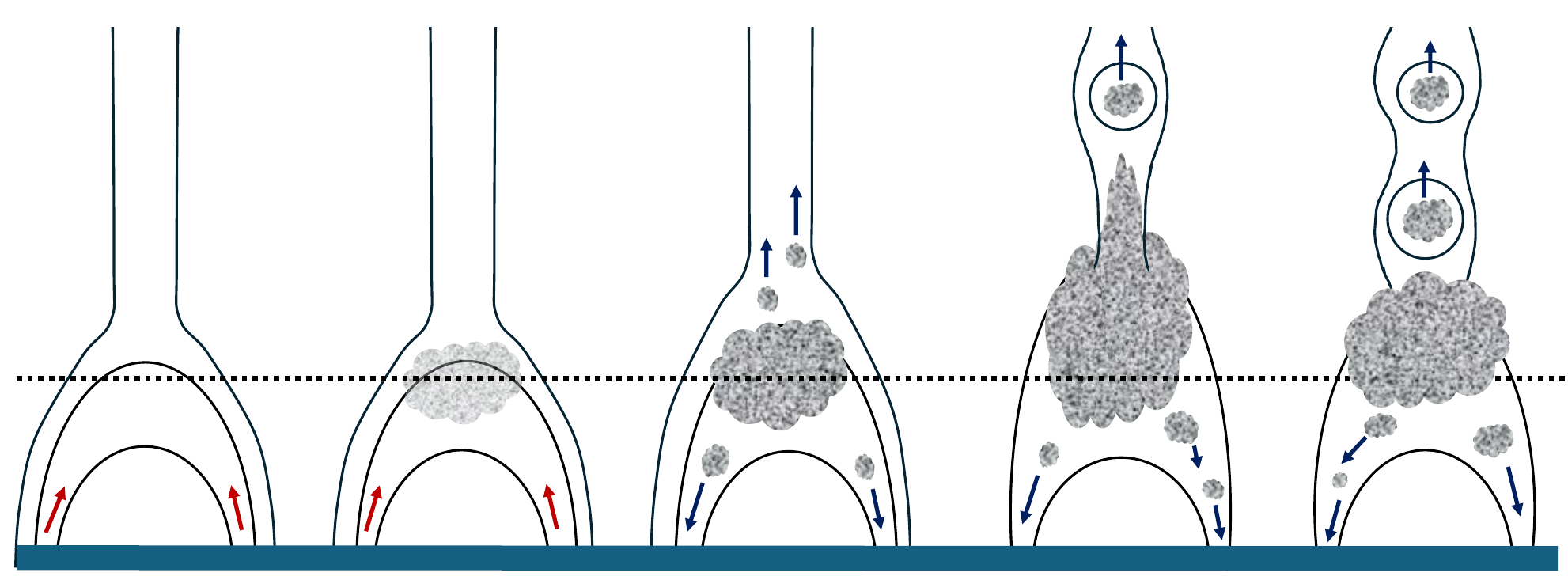}
	\caption[]{Development of cyclic ejection of gas clouds. From left to right, the five stages are: (i) heating concentrated at low heights evaporates plasma from the chromosphere (shown as the lower blue horizontal line); (ii) at larger heights there is insufficient heating to prevent cooling;  (iii) condensations form and rain out of the cloud either downwards (below the co-rotation radius which is shown as the black horizontal dotted line) or outwards (above co-rotation); (iv) the tearing mode produces plasmoids which (v) are centrifugally ejected.\label{fig:cartoon}}
\end{figure*}

Additional lines of evidence have come from detailed photometric and spectroscopic studies of young clusters, including Upper Scorpius and the Pleiades \citep{2016ApJ...816...69A,Rebull2016, Rebull2018, Stauffer2021, Gallet2017, Somers2020}. These allow us to place the ``scallop-shell'' stars within a wide spectrum of stars that show periodic and stochastic photometric dips. There is a range of behaviours, both in the depth and lifetime of the dips and the existence (or absence) of an accompanying IR signature consistent with a disk. \citet{2019MNRAS.484.4260S} suggest an evolutionary sequence that spans both the ``dipper'' stars that appear to possess full disks \citep{2016ApJ...816...69A}, through to a regime where the disk has fragmented to leave only clumps that produce irregular dips. Scallop-shell stars have regular, shallow, photometric dips and no direct indication of disks, and so might form an even later stage. Their incidence rate and the fact that they are found preferentially among the most rapidly rotating low-mass members of these studies suggest that rotationally driven magnetic activity, combined with the presence of dust, is a key factor in their formation. Moreover, the abrupt state changes observed in some systems, frequently following stellar flares, highlight the dynamic interplay between magnetic reconnection events, mass ejection, and dust cloud stability.

A complementary avenue of investigation has emerged from the study of radio emission from scallop-shell stars. \citet{Kaur2024, Kaur2025} reported the detection of polarized radio signals from two different scallop-shell M dwarfs, identifying both persistent gyrosynchrotron emission and episodic, highly polarized bursts likely due to an electron cyclotron maser. The occurrence of polarisation reversals in these signals provides compelling evidence for the presence of large-scale magnetic structures modulating the emission and potentially interacting with the particulate clouds responsible for optical variability. In addition, recent modelling of the radio corona of the rapidly rotating K-dwarf AB Dor  demonstrates how large-scale closed magnetic loops and centrifugally supported prominences can shape free–free radio emission \citep{Brasseur2024}, while the ejection of these prominences may drive bursts of reconnection and particle acceleration, producing electron cyclotron mesa instability (ECMI) emission \citep{10.1093/mnras/stag042}. Although AB Dor is not a confirmed scallop-shell star, the physical picture it presents—an extended, structured magnetosphere capable of hosting, and periodically ejecting, dense co-rotating material—offers a useful analogue for interpreting the radio behaviour of young M-dwarfs. This dual observational signature, manifesting in both optical and radio regimes, underscores the magnetospheric origin of the scallop-shell phenomenon. 

In parallel to observational studies, theoretical work has advanced the case for magnetically confined dust clouds as the principal drivers of scallop-shell variability. This builds upon prior theoretical frameworks established for clouds of cool gas suspended within the coronae of rapidly-rotating low-mass stars \citep{Ferreira2000, Jardine2001, Jardine2020, Villarreal2021} and extends them to encompass dust-rich environments. Such cool prominence structures have been studied in the solar corona and have a rich history of literature on the subject. Recent numerical studies include \cite{Xia2016, Jercic2023, Lu2024, Johnston2025}. See the living review by \cite{Keppens2025} and references therein for greater details.

A detailed treatment of dust grain dynamics in stellar magnetospheres has been provided by \citet{Sanderson2023}. This work demonstrates that micron-sized grains (sourced either from the star or externally), can become collisionally charged in the hot coronal plasma and migrate along field lines to stable equilibrium points near or just beyond the co-rotation radius. The longevity and mass distribution of such dust clouds, as modelled from observed systems, are consistent with the amplitude and duration of the photometric dips recorded in Kepler and TESS data. 

The potential connection between scallop-shell dips and disintegrating planetary bodies has also gained traction. Studies of systems include a range of planetary masses from hot Jupiters \citep{Daley-Yates2018, Daley-Yates2019} to lower mass planets. While currently known disintegrating planets such as KIC 1255b \citep{Rappaport2012}, K2-22b \citep{2025ApJ...987L...6T}, and BD+05b \citep{2025ApJ...984L...3H} are all seen around old stars with weak magnetic fields, the dust dynamics seen in the planetary outflows exist on a continuum with the strong magnetic field limit seen in the scallop-shell stars.

The evolution of dusty tails around disintegrating rocky bodies \citep{Debes2016, vanLieshout2018} have shown that sublimation and tidal disruption can generate highly-variable, asymmetric transits. See \cite{Vidotto2025} for a recent review on the subject of star--planet interaction. 

The morphological similarity between these features and scallop-shell dips suggests that planetary debris may serve as a source of the dust trapped in stellar magnetospheres. {The dust may also originate from the remnant of a protoplanetary disks. However, such disks are usually observed as a infrared excess. This is not seen in observations of scallop-shell stars \citep{Stauffer2017}, and indicates that their primordial disks should have dissipated. This however, it does not rule out residual debris or planetesimal populations that may persist and contribute to observed variability.}

A recent breakthrough has come from time-series spectroscopic and photometric observations of TIC 141146667, a rapidly-rotating young M-dwarf (P$_{\rm rot}$=3.9hr, M$_\star = 0.22M_\odot$, R$_\star = 0.42R_\odot$). These observations have provided clear evidence that the material responsible for the photometric dips is co-spatial with the clouds of cool gas that give rise to the H$_{\alpha}$ dips\footnote{It should be noted that \cite{Bouma2025} showed that for TIC 141146667, transits of the inner H$_{\alpha}$ clump only partially overlap the complex photometric modulation.}. This has brought into sharper focus the question of the relationship between the dynamics of the gas and dust. The gas clouds are observed to undergo ejection events every day or so \citep{CollierCameron1989}. As noted by \citet{Sanderson2023} this presents a problem if the dust is entrained in the coronal gas, since these ejection events would remove dust far more rapidly than the lifetime of the dips suggests. 

Modelling of the gas behaviour \citep{Daley-Yates2024} has demonstrated that the periods of these ejection events are similar to what is observed. Crucially, however, it has also revealed that only a fraction of the mass of the gas cloud is removed in each ejection and a large reservoir of gas remains around the co-rotation radius (see Fig. \ref{fig:cartoon}). The possibility remains, then, that the dust may survive trapped in the gas clouds for timescales that are much longer than the ejection events. 

Taken together, these observational and theoretical threads converge on a model in which scallop-shell stars represent a nexus of rapid rotation, strong magnetic fields, and dust-bearing environments shaped by in-situ dust formation, ongoing planetary disintegration, collisional cascades or comet injection. 

In this paper we model the corona of a rapidly-rotating M-dwarf into which dust-like particles have been injected. We aim to determine the timescale over which these particles might remain trapped and compare it to the timescales for the gas ejection events. By comparing the dust lifetime with the observed lifetimes of the dips, we aim to determine if dust injected into an M-dwarf magnetosphere can survive for long enough to explain the observed photometric dips.

In Section \ref{sec:modelling} we provide an overview of the slingshot prominence process and describe how we extend our previous modelling to describe dust-bearing prominences. We will present our results in Section \ref{sec:results} and discuss their implications for the scallop-shell phenomenon in Section \ref{sec:discussion}.

\begin{figure*}
	\centering
	\includegraphics[width=0.99\textwidth,trim={0cm, 0cm, 0cm, 0cm},clip]{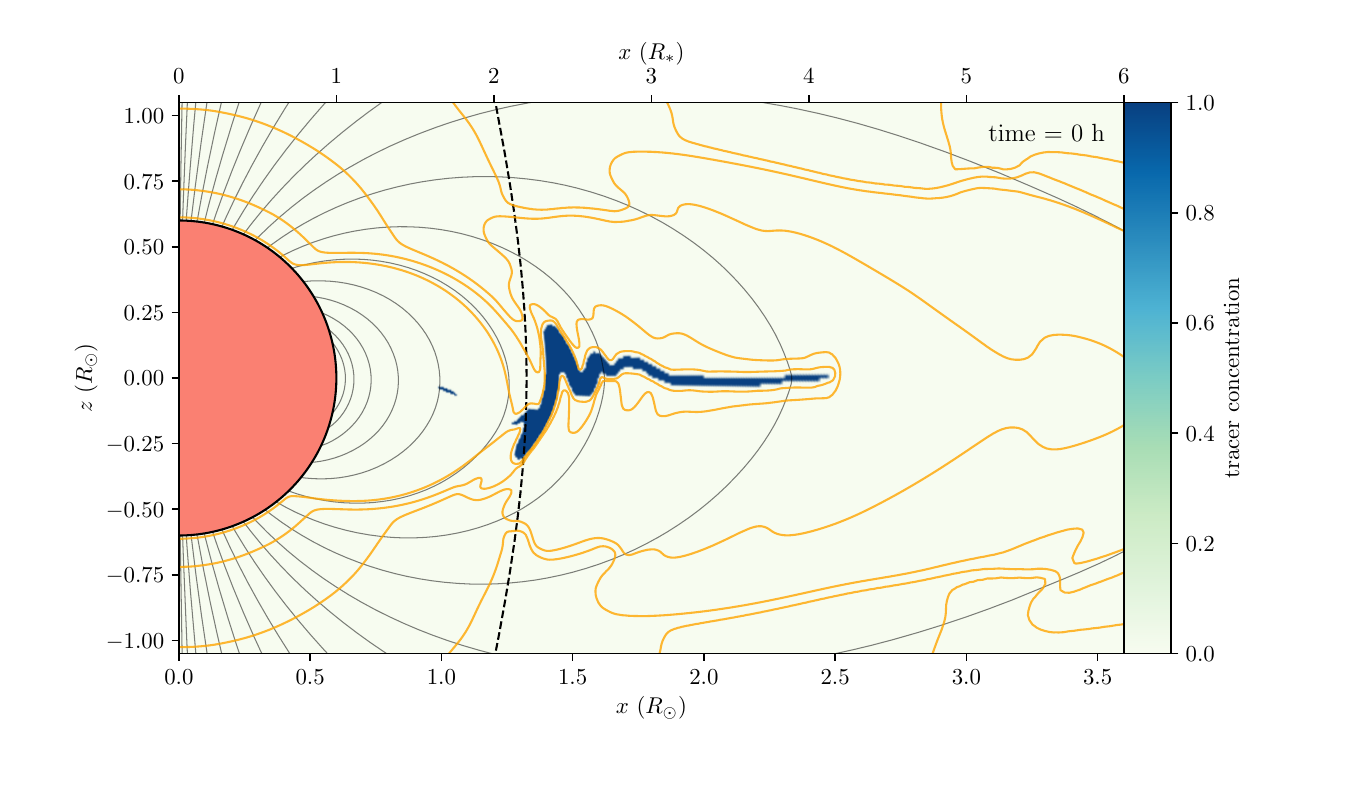}
    \includegraphics[width=0.99\textwidth,trim={0cm, 0cm, 0cm, 0cm},clip]{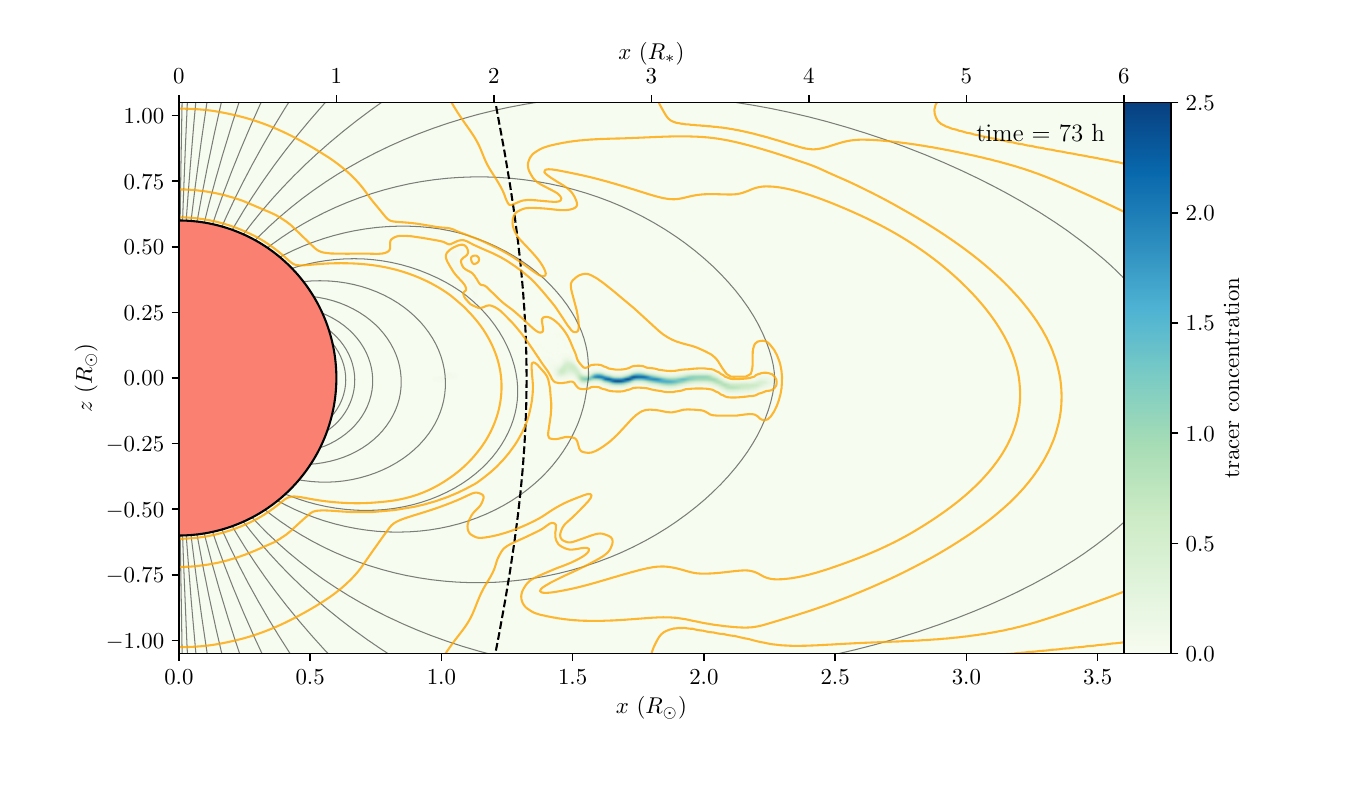}
	\caption[]{Two dimensional plots of the dust structure around our simulated star. The top is the initial state with uniform tracers in the prominence region. Bottom is the final state. The orange contours depict the density distribution and are equidistant in log space. The black lines show the magnetic field geometry and the colour map shows the tracer, and therefore dust, concentration. The dust bearing prominence is located between 2.2 and 3.8 $R_{\ast}$ above the stellar surface. This places it beyond co-rotation, illustrated by the dashed black line, where centrifugal acceleration is balanced by magnetic tension. This balance confines the gas and dust in a stable equilibrium. Periodic breakouts from the apex of the prominence, known as slingshots, remove the dust from the prominence, while gas is replenished from the evaporating chromosphere at the lower boundary. The top axes show the scale in stellar radii and the left and bottom show the scale in solar radii. An animated version of this figure is available online. \label{fig:tracer_location}}
\end{figure*}

\begin{figure}
	\centering
	\includegraphics[width=0.49\textwidth,trim={0cm, 0cm, 0cm, 0cm},clip]{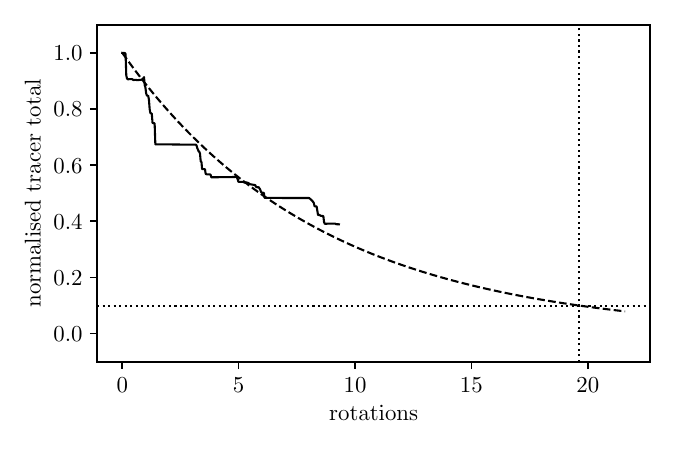}
	\caption[]{Tracer decay. Here we integrate the tracer concentration over the whole simulation grid, for all outputs. We then normalise each output to the total integrated tracers from the first time step. This way we can see how the action of cyclic slingshot ejections remove tracers, and therefore dust from the prominence (solid line). We then fit a exponential (dashed line) to this decay and work out the dust prominence half-life. From the fit, the half life of tracers on the grid is 45.8 hr (5.9 rotations). This means that after 152.0 hr (19.6 rotations), the tracer concentration has dropped to 10\% of it's original value. We highlight this \textit{tenth-life} on the plot with the coinciding dotted lines.\label{fig:tracer_decay}}
\end{figure}

\begin{figure*}
	\centering
	\includegraphics[width=0.99\textwidth,trim={0cm, 0cm, 0cm, 0cm},clip]{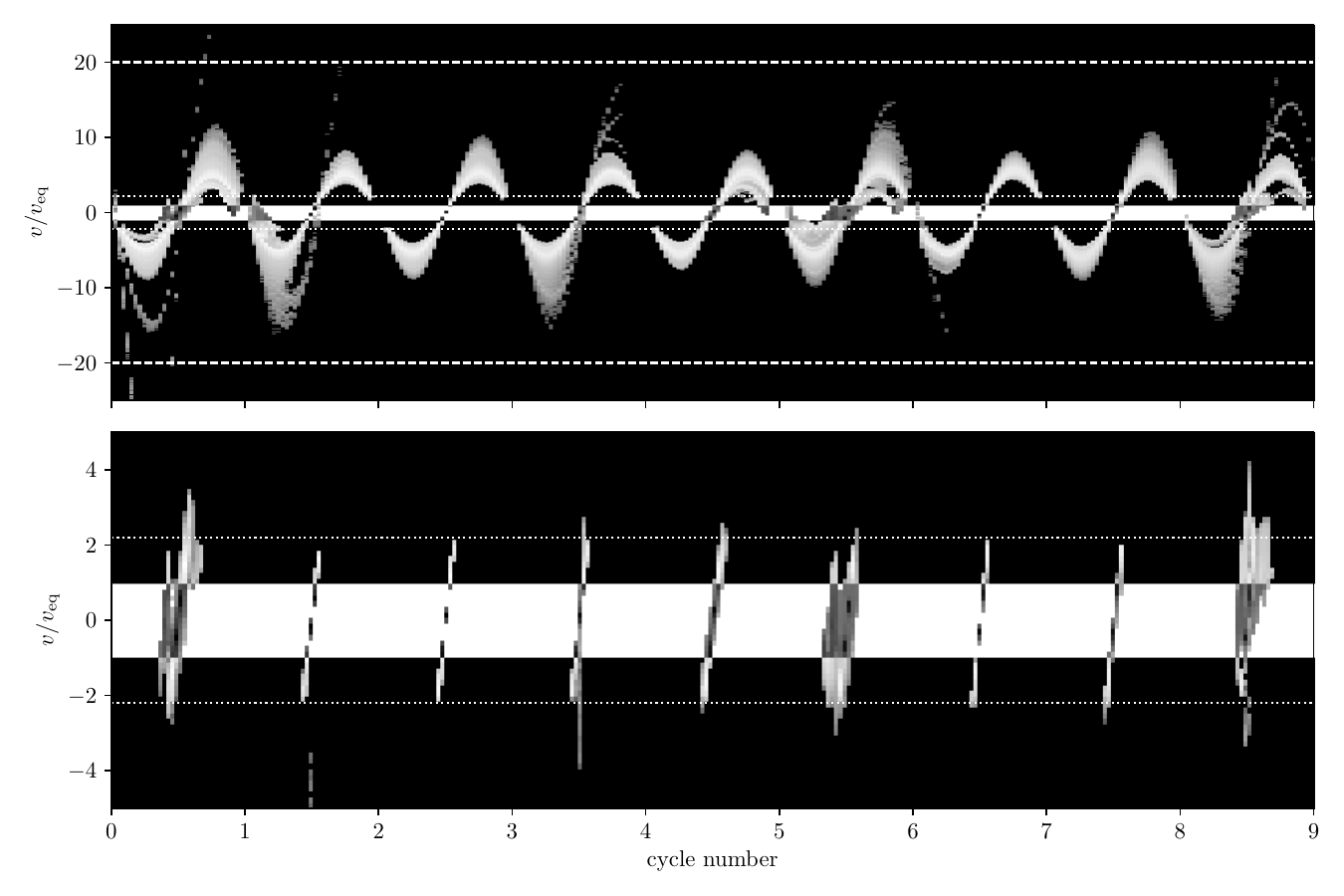}
	\caption[]{Dynamic spectrum of the prominence gas, which is illustrative of mock H$_{\alpha}$ tracks. The colour map is normalised between zero and one, as the absolute value is not important, {the rate at which the feature travels through the line centre gives the distance of from the rotational axis. The faster the feature crosses, the further it is from the axis} (see \cite{Jardine2020} for examples of this method). Each cell in the simulation is binned according to its velocity after projection onto the observers line of sight. Top: binned velocity from all cells with tracers above 1\% of the initial concentration. The dashed line corresponds to the ridged rotational velocity of at the average Alfvén surface radius ($20 \ R_{\ast}$). Bottom: binned velocity from all cells that are both above 1\% of the initial concentration and transit the stellar disk from the observers point of view. The dotted line corresponds to the rigid rotational velocity of the co-rotation surface ($2.21 \ R_{\ast}$). The white core represents the rigid body rotational velocity of the stellar surface.\label{fig:dynamic_spectrum}}
\end{figure*}

\begin{figure}
	\centering
	\includegraphics[width=0.49\textwidth,trim={0cm, 0cm, 0cm, 0cm},clip]{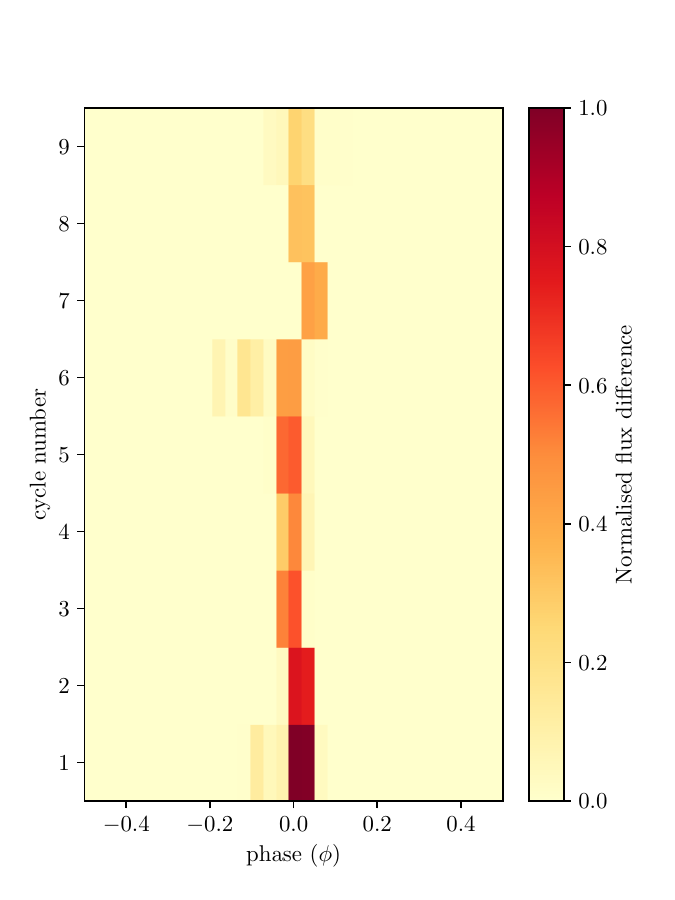}
	\caption[]{River plot from the dynamic spectrum Fig. \ref{fig:dynamic_spectrum}. A single feature persists at phase angle $\phi=0$. Our simulated 2D dipole has been cylindrically rotated about the $z$-axis to form the pseudo-3D signal. At $t=0$ the prominence, star and observer all line up. This is why we always have a single persist feature at a phase angle $\phi=0$. the colour map is normalised to the peak initial value in the first cycle. The decay of the feature with cycle number is due to the leaking out of the dust from the prominence with successive slingshots or with coronal rain and has reduces to approximately half its initial value over the 9 cycles of the simulation. This is consistent with some of the shorter dip lifetimes reported by \cite{Bouma2024}.\label{fig:river_plot}}
\end{figure}

\section{Modelling}
\label{sec:modelling}

\begin{table}
	\centering
	\caption[]{{Representative stellar parameters}.
	\label{tab:parameters}}
	\begin{tabular}{cc}
		\hline
		Parameter & Value \\
		\hline
 		Radius & 0.6 $R_{\sun}$ \\
		Mass & 0.3 $M_{\sun}$ \\
		Period & 0.32 day \\
        Co-rotation radius & 2.21 $R_{\ast}$ \\
        Polar magnetic field strength & 693 G \\
		\hline
	\end{tabular}
\end{table}

For the equations solved in our simulation including the form of the magnetohydrodynamic (MHD) equations and source terms, see our past publications \cite{Daley-Yates2023} and \cite{Daley-Yates2024} and references therein. These works also describe the simulation's magnetic field geometry, together with our treatment of the chromosphere, transition region and corona. This setup is implemented with the public MHD code \textit{MPI-AMRVAC} \citep{Xia2018, Keppens2023}.

{We normalise length, time, and mass in the simulation so that the results can be rescaled to different stellar parameters, following an approach similar to \cite{Raives2023}. For the present study, we adopt a normalisation corresponding to a rapidly rotating M-dwarf star, representative of those exhibiting scallop-shell light curves. The choice of parameters is observationally motivated. The stellar mass and radius are chosen such that they are representative of the scallop-shell star population as presented by \cite{Bouma2024}. Once the mass radius is fixed, we select a rotation period characteristic of fast rotators with this mass-radius combination. The surface magnetic field strength is then determined from this rotation rate using established empirical scaling relations (see for example \citealt{Vidotto2014, Reiners2022}). In this way, all normalisation parameters are physically linked rather than independently specified. The specific values adopted are listed in Tab. \ref{tab:parameters}.}

\subsection{The slingshot prominence phenomena}
\label{sec:slingshots}
To understand how prominence gas can house dust in a coronal equilibrium point and also carry it away by centrifugal ejection, we need to understand the notion of a slingshot prominence. In our previous work \citep{Daley-Yates2024} we demonstrated that under the assumptions of MHD, a rapidly rotating magnetically active star has a magnetosphere where gravity, magnetic tension and centrifugal acceleration can balance, resulting in mechanically stable points in the corona. At these stable points, gas that has cooled via thermal instability, accumulates  until centrifugal acceleration and gas pressure overcomes magnetic tension and prominence material is ejected as plasmoids. This process is cyclic and is summarised in the following steps (see Fig. \ref{fig:cartoon}):
\begin{enumerate}
    \item The chromospheric evaporation stage.
    \item The cooling stage.
    \item Condensation and draining {\it below}, and centrifugal breakout {\it above} the Kepler co-rotation radius.
    \item The tearing instability produces plasmoids.
    \item The plasmoids coalescence beyond the Alfv\'{e}n surface.
\end{enumerate}
The resultant gas and dust dynamics we present in this paper should be understood in the context of this cyclic behaviour.

The magnetospheric dynamics of massive stars share analogies with slingshot prominences, in that large-scale magnetic fields can support mechanically stable coronal plasma against gravity and centrifugal forces \citep{ud-doula2002}. In massive stars, this balance arises from the magnetic channelling of radiatively driven winds, with opposing flows colliding near the magnetic equator to produce hot, X-ray–emitting wind collision regions \citep{ud-doula2006, ud-Doula2008}, sustained by continuous wind feeding and exhibiting rotational modulation \citep{ud-Doula2016, Daley-Yates2019a}. By contrast, slingshot prominences in cool stars are mass-loaded by chromospheric evaporation and shaped by thermal instability, leading to the condensation of cool material and episodic centrifugal ejection. See \cite{Villarreal2018} for a comparison of the rotational and magnetic proprieties of massive and cool stars. Existing models of massive-star winds and magnetospheres do not exclude the possibility of prominence formation, however they do typically adopt isothermal treatments or impose temperature floors near the stellar effective temperature, suppressing cool condensations. Their absence may therefore reflect modelling assumptions rather than a fundamental physical difference. This comparison is intended to provide context rather than to suggest a unified treatment across stellar types. This paper focuses specifically on the cool-star slingshot prominence regime.

\subsection{Dust as a passive scalar}
\label{sec:tracer_dust}

The dust thought to be responsible for the scallop-shell phenomena is collisionally charged. The simulation framework we use, \textit{MPI-AMRVAC}, does not currently have the ability to model charged dust. We therefore do not treat the dust directly, instead we use a passive advected scalar field, otherwise known as a tracer. Under this assumption, the dust is in the fully coupled regime, where any motion of the gas is instantly mirrored in the dust. \citet{Sanderson2023} considered diffusion timescale of dust from a prominence and \cite{Zhan2019} studied sublimation timescales of dust around M-dwarfs. Both timescales were found to be many orders of magnitude longer than time scale of observed signal decay, which have been found to persist over months, and spanning tens to thousands of rotation cycles \citep{Bouma2024}. This means that the reduction in dust concentration by diffusion and sublimation is negligible compared to the slingshot gas ejection time scale which we found to be on the same order of the rotational period \cite{Daley-Yates2024}. All this means that any estimate of dust evacuation timescales we make in this study are lower limits. Dust evacuation timescales could be longer if the fully coupled assumption is relaxed.

{The role of the stellar wind in the dust dynamics also depends on the magnetic topology. The prominence formation region lies within the so-called \textit{dead zone}, where magnetic field lines are closed and plasma flows are relatively weak compared to the free-streaming stellar wind along open field lines. Under these conditions, the dynamics are dominated by the confined coronal plasma. Following a slingshot ejection, however, the gas and any entrained dust transitions onto open field lines beyond the Alfv\'{e}n radius and into the stellar wind outflow. Consequently, while the stellar wind influences the subsequent transport of ejected material, it does not significantly affect the dust evolution at the prominence location prior to ejection.}

As the simulation evolves the tracers will move from their initial location within the prominence, either by advection or by numerical diffusion. As such, even at early times, there is a non-zero concentration of tracers between the prominence and the stellar surface as well as at higher and lower latitudes out at the prominence orbit. To avoid binning cells with very low tracer concentrations, we set a lower cut-off of tracer concentration to 1\% of the initial value. This ensures we only consider cells with tracers within the prominence cells where dust will not survive.

\section{Results}
\label{sec:results}

\subsection{Morphology and evolution of the dust-bearing prominence}

Fig.~\ref{fig:tracer_location} shows the evolution of the dust and gas in the simulated magnetosphere. Initially, the tracer---representing the dust concentration---is uniformly distributed within the prominence region. As the system evolves, prominence gas and dust evolve along closed field lines between $2.2$ and $3.8 \, R_{\ast}$ above the stellar surface and beyond the co-rotation radius. Here centrifugal acceleration is balanced by magnetic tension, providing a stable environment for material accumulation. The density contours, spaced logarithmically, highlight compression in the prominence core, while the field-line geometry demonstrates how confinement is maintained throughout most of the cycle. At the prominence apex, the balance occasionally fails, producing brief breakout events that expel gas and dust from the prominence. These events are followed by renewed condensation as chromospheric evaporation continuously replenishes the prominence from the lower foot-points. This cycle of confinement, breakout, and continuous replenishment defines the quasi-steady evolution of the gas reservoir that the dust resides in.

\subsection{Tracer depletion and prominence lifetime}

To quantify the removal of dust from the system, we integrated the total tracer content over the grid at each output time and normalised to its initial value. The resulting decay curve in Fig.~\ref{fig:tracer_decay} shows a shallow reduction in tracer concentration punctuated by abrupt steps coincident with individual breakout episodes. Fitting an exponential to the envelope of the decay yields a characteristic timescale that describes the depletion of the dust from the prominence reservoir. From this fit, a half-life of approximately $45.8$~hr ($5.9$ rotations) is obtained, corresponding to a reduction to one tenth of the initial tracer content after $152.0$~hr (19.6 rotations). This timescale reflects the combined effects of recurrent centrifugal ejections and continuous surface feeding, and provides a measure of the lifetime of the dust component of the prominence. The exponential behaviour implies that, despite their impulsive nature, individual breakouts collectively remove material at an approximately steady fractional rate.

\subsection{Kinematic synthetic spectra}

The dynamics of the tracer-bearing gas were examined by constructing velocity--phase maps analogous to observational dynamic spectra. These diagnostics are derived from our two-dimensional simulation by varying the observer’s viewing angle relative to the simulation plane, rather than by assuming cylindrical symmetry of the flow. Specifically, the simulation domain is not extruded azimuthally about the $z$-axis. Instead, the observer is repositioned at successive rotational phases, and for each phase the line-of-sight (LOS) velocity of the plasma is computed in the observer’s frame. This procedure preserves the intrinsic two-dimensional structure of the magnetic field and prominence while allowing its kinematic signature to be sampled as the star rotates.

Within this framework, the simulated magnetic structure and associated prominence are treated as the centre of a single, large-scale active region. This assumption implies that only one dominant absorbing structure contributes to the synthetic spectra. If multiple active regions were present at different longitudes, each hosting a dust-bearing prominence, the resulting spectra would exhibit multiple discrete features per rotation cycle. Since the aim of this work is to isolate the observational signature and decay behaviour of a single dust-bearing prominence, this simplified geometry is sufficient for the present analysis.

For each simulation snapshot, cells were binned by their projected LOS velocity. When considering all cells above the tracer concentration threshold (see Section~\ref{sec:tracer_dust}), the upper panel of Fig.~\ref{fig:dynamic_spectrum} shows a broad velocity distribution bounded by the rotational velocity at the average Alfv\'en surface ($20\,R_{\ast}$). Restricting the selection to cells that both exceed the tracer threshold and intersect the stellar disk isolates the component that contributes to absorption signatures. This produces a narrower velocity feature that traces the rigid rotation of the co-rotation surface at $2.21\,R_{\ast}$ and reveals a single coherent structure per stellar rotation. The white central region corresponds to the projected velocity of the stellar surface, against which the prominence feature migrates smoothly with phase.

The resulting velocity--phase behaviour is consistent with the rotation of a confined condensation anchored to a fixed magnetic longitude. Although this approach does not capture interactions between multiple prominences or longitudinal asymmetries, it provides a controlled and physically motivated framework for connecting prominence dynamics, dust depletion, and the evolution of individual features in synthetic spectra.

\subsection{Phase-folded persistence and decay}

The dynamic spectrum was folded over successive rotation cycles to form the river plot shown in Fig.~\ref{fig:river_plot}. A single, phase-stationary feature persists at $\phi = 0$, the point where the prominence, stellar disk, and observer are aligned. Because the underlying two-dimensional simulation plane is rotated cylindrically about the symmetry axis, the feature remains fixed in phase. Its intensity, normalised to the peak of the first cycle, declines steadily with time. After nine cycles the amplitude has fallen to roughly half its initial value, consistent with the tracer decay curve and indicating that the signal is controlled by the gradual leakage of dust through centrifugal breakout and small-scale draining. The stability of the phase position, despite the declining strength, confirms that the feature originates from a single long-lived prominence anchored at a fixed longitude.

\section{Discussion}
\label{sec:discussion}

Together, the morphological, temporal, and kinematic behaviour outlined above describe a coherent sequence. Cold gas accumulates in magnetically closed loops beyond co-rotation, forming a quasi-static reservoir that periodically releases material through centrifugal ejections. Dust which has been introduced to the prominence is entrained in the gas and the dust dynamics depend upon the forces acting on the dust grains. 

\citet{Sanderson2023} demonstrated analytically that charged dust grains in stellar magnetospheres can become collisionally charged in the hot corona and channelled along magnetic field lines into stable points. The force on a dust grain in the presence of a magnetic field is a combination of the Lorentz force and drag between the gas and dust.
\begin{equation}
    \boldsymbol{F}_{\mathrm{total}} = q_{\mathrm{d}} \left( \boldsymbol{v_{\mathrm{d}}} \times \boldsymbol{B} \right) + m_{\mathrm{p}} n_{\mathrm{p}} \sigma_{\mathrm{col}} \boldsymbol{v}_{\mathrm{rel}} + \pmb{\textsl{g}}_{\rm{eff}}
    \label{eq:dust_force}
\end{equation}
where $q_{\mathrm{d}}$ is the charge on the dust grain, $\boldsymbol{v}_{\mathrm{d}}$ is its velocity and $\boldsymbol{B}$ the  magnetic field.
$m_{\mathrm{p}}$ and $n_{\mathrm{p}}$ are the mass and number density of protons that comprise the gas and $\sigma_{\mathrm{col}}$ is the effective cross-section of a dust grain. $\boldsymbol{v}_{\mathrm{rel}}$ is the relative velocity between the dust and gas. $\pmb{\textsl{g}}_{\rm{eff}}$ is the effective gravity and accounts for both Coriolis and centrifugal forces {(see \citealt{Waugh2022} for a complete discussion).} The first term on the right hand side is the Lorentz force (assuming no electric fields) and the second is the drag force. By inspection we see that {when the dust is at a stationary} point in the prominence, where the gas is also stationary, the total force on the dust grain is zero. Over time, centrifugal acceleration of the prominence gas leads to the gas pressure building at the apex of the closed field lines. Once the gas pressure exceeds magnetic tension, centrifugal breakout occurs and some of the prominence gas is ejected. For this to happen the gas must move along the field lines as they elongate. This would make the drag term in equation (\ref{eq:dust_force}) non-zero and force the dust to the apex of the field lines. Reconnection occurs and the slingshot is released into the stellar wind as a magnetically-isolated plasmoid. Any dust that accompanies the gas will be trapped in the plasmoid and removed.

This force balance implies that, while confined, the dust remains dynamically coupled to the gas and follows the same magnetic topology, justifying the use of a passive tracer to model dust transport within the prominence and its subsequent removal during centrifugal breakout. In this sense, our treatment represents an upper limit on the degree of gas--dust coupling and therefore an upper limit on the efficiency with which dust is removed from magnetic confinement. Any more complete treatment that relaxes the assumption of perfect coupling---for example by explicitly evolving grain charge, inertia, or differential drift---would reduce the effectiveness of the drag and Lorentz terms in equation~(\ref{eq:dust_force}), allowing dust to remain confined for longer than in the present model. The decay timescales inferred here should therefore be interpreted as lower limits on the true dust residence time within the prominence. Longer-lived confinement is more consistent with the persistence of dust-associated features reported by \citet{Bouma2024}, suggesting that the prominence lifetimes inferred from our tracer decay provide a conservative bound rather than a definitive estimate.

Each breakout produces a sharp decline in the integrated tracer (Fig.~\ref{fig:tracer_decay}) and a transient velocity feature in the dynamic spectrum (Fig.~\ref{fig:dynamic_spectrum}). Phase folding reveals that these features remain phase-locked even as their strength diminishes (Fig.~\ref{fig:river_plot}), reflecting the longevity of the magnetic structure confining the dust. The half-life derived from the tracer decay and the cycle-to-cycle fading of the spectral feature therefore represent the same physical process: the progressive exhaustion of the dust from the prominence by repeated slingshot events.

An additional feature of the dynamic spectrum in Fig.~\ref{fig:dynamic_spectrum} is that it shows secular variation. The prominence evolves throughout the simulation, so for each transit of the star, the prominence morphology is slightly different. We can see this in the lower plot, where there is cycle to cycle variation in the feature and therefore in the LOS velocity component of the dust. We can interpret this as both the coronal rain and slingshot behaviour, as both require motion either away from (in the case of rain) and towards (in the case of slingshots) the observer relative to the bulk of the prominence material. This secular variation is reported in the observation of \cite{Bouma2024} and \cite{Bouma2025} and is further evidence that the scallop-shell phenomenon is due to the presence of centrifugally supported prominences with periodic coronal rain and slingshot breakouts.

\cite{Sanderson2023} assumed that with each slingshot the entire prominence structure would have to be expelled from the corona. For fast rotating stars exhibiting the scallop-shell phenomena, this happens multiple times in a stellar rotational cycle, while the dust signature survives for many hundreds of cycles. This apparently presents a problem. However, what we find in our simulation \citep{Daley-Yates2024} is that it is only the upper part of the prominence, which has extended furthest from the star, that is undergoing slingshots. The rest remains near co-rotation and acts as a reservoir, continuously replenished by chromospheric evaporation, and feeding both the coronal rain and slingshots that periodically drain the dust from the prominence. This picture allows for the co-location of the prominence gas and the dust and was not known by \cite{Sanderson2023}.
 
We find that for our simulated star, the tracer decays by an order of magnitude over 20 stellar rotations. This is reflected in our synthetic river plot in Fig. \ref{fig:river_plot}, where the signal intensity diminishes steadily over the nine simulated rotation cycles. This reproduces one of the types of river plot features which \cite{Bouma2024} observed while analysing the light curves and river plots of the M-dwarf star LP 12-502. \cite{Bouma2024} describes features that are phase locked, persistent over tens to thousands of cycles and can show either gradual decay or sudden state changes where a feature can disappear in less than one rotational cycle. Motivated by our simulation results, we believe the river plot features undergoing gradual decay, on the order of tens of cycles, are due to stable prominences that exhibit the slingshot phenomenon. 

River plot features that persist and show no decay may still be due to dust trapped in prominences that are not undergoing slingshots. These types of prominences on the Sun are termed quiescent prominences and indeed can persist in the solar atmosphere for tens of months. These quiescent prominences must necessarily be suspended lower in the star's corona, below co-rotation, in order to avoid undergoing slingshotting and decay. 

The width of a feature in a river plot indicates the stellar rotation phases during which the absorbing body transits the stellar disk. The thicker the feature, the smaller the orbital radius of the transiting object. This means that if persistent features that do not decay are due to quiescent prominences below co-rotation, their river plot features should be thicker than those at co-rotation undergoing slingshots.

Supporting a prominence without centrifugal acceleration requires that the star has a magnetic topology containing magnetic dips \citep{Aulanier1998, Aulanier1998A, Ballegooijen2004, Yingna2012, Chen2025}. These dips allow cool gas to pool in an equilibrium point that is just a balance between gravity and magnetic tension. Despite these prominences being stable in the Sun's corona for many months they can undergo rapid ejection when the magnetic field topology reconfigures. These same similar processes occur on other M-dwarf stars where flaring activity is observed in combination with H$_{\alpha}$ excess indicative of a stellar mass ejection \cite{Vida2019}. 

\section{Conclusions}

Our simulation shows that dust entrained within magnetically confined prominences can remain trapped for timescales comparable to those inferred from scallop-shell variability. The tracer field, used to represent fully coupled dust, decays exponentially with a half-life of roughly $46\,\mathrm{hr}$, as successive slingshot ejections remove material from the prominence apex. These timescales are specific to the stellar parameters adopted here and are expected to vary systematically with rotation rate, magnetic field strength, and stellar mass. Because the dust is treated as fully coupled to the gas, these decay times should be interpreted as lower limits on the true dust residence time, with more realistic treatments expected to prolong confinement. Despite these losses, the bulk of the gas and dust is retained near the co-rotation radius, where the gas (but not the dust) is replenished continuously by chromospheric evaporation. This quasi-steady balance produces long-lived prominences that reproduce the phase-locked, slowly fading features seen in observations.

The persistence of these prominences implies that charged dust grains can survive repeated breakout events without total loss, rapid sublimation or diffusion. The morphology and temporal evolution of the synthetic spectra closely match the river-plot features of scallop-shell stars, supporting the interpretation that these light-curve modulations arise from dust trapped in coronal magnetic loops undergoing cyclic centrifugal ejections.  

We can interpret the variety of river-plot morphologies observed in LP~12-502 within this framework. Features that persist with no measurable decay correspond to quiescent prominences held below co-rotation, supported by magnetic dips where gravity balances magnetic tension. These structures remain stable for many rotations and produce the broadest features in the river plots. Narrower, gradually fading features arise from prominences located near the co-rotation radius that undergo periodic centrifugal breakouts---the slingshot events---which steadily drain dust from the system. Finally, features that vanish abruptly in less than a single rotation are best explained by flare-triggered reconnection and the rapid expulsion of the prominence. The relative width of each feature provides an additional diagnostic: thicker, more extended bands imply lower-altitude structures, while thinner, sharper features trace material closer to or beyond co-rotation. Together, these classifications link the observed photometric behaviour of scallop-shell stars to distinct regimes of prominence stability in their magnetised coronae.

A positive correlation between the width of river plot features and their lifetimes would support this hypothesis. If however a {\it lack} of correlation can be demonstrated, then the persistent features that show no measurable decay may instead represent locations in the stellar magnetosphere where there is continuous injection of dust. This could co-exist with smaller, transient inflows at other locations.

While a low rate of comet delivery may provide only intermittent injection of dust, a more continuous injection could be supplied by, for example, the footpoint of a quasi-stable accretion column from a disintegrating planetary body. In both cases, dust would be supplied only to discrete locations, so not every stable location in the corona that supports cool gas would also contain dust. Continuous delivery at {\it all} sites of condensed gas would be provided if the dust were able to form directly out of the cool gas. In this scenario, the formation of a gas condensation would lead to the formation of dust.  Modelling this process (which would require pockets of very low temperatures within the cool gas clouds) is however beyond the scope of this paper. 

The timescale on which some river plot features decay provides valuable information about the nature of the dust delivery, as if sets a lower limit to the interval between discrete injections of dust. For example, an interval of three days between injections (typical of the rate of present-day Sun-grazing comets) would be sufficiently long to allow our modelled features to decay.

We conclude that magnetically supported dust-gas clouds can exist stably in the coronae of young, rapidly rotating M-dwarfs. Their evolution is governed by the interplay between centrifugal breakout, magnetic tension, and continuous mass loading from the chromosphere. This framework provides a natural explanation for the longevity and coherence of scallop-shell photometric dips and links them directly to the dynamics of slingshot prominences.

\section*{Data availability}

The data presented in this article will be available from the University of St Andrews research repository.

\section*{Acknowledgements}

The authors thank the reviewer for their helpful comments and suggestions; which improved the quality and content of the publication. 
SD-Y and MJ acknowledge support from STFC consolidated grant number ST/R000824/1. This work was performed using the DiRAC Data Intensive service at Leicester, operated by the University of Leicester IT Services, which forms part of the STFC DiRAC HPC Facility (\url{www.dirac.ac.uk}). The equipment was funded by BEIS capital funding via STFC capital grants ST/K000373/1 and ST/R002363/1 and STFC DiRAC Operations grant ST/R001014/1. DiRAC is part of the National e-Infrastructure. For the purpose of open access, the authors have applied a Creative Commons Attribution (CC BY) licence to any Author Accepted Manuscript version arising.

\section*{ORICD IDs}

Simon Daley-Yates \orcidlink{0000-0002-0461-3029} \url{https://orcid.org/0000-0002-0461-3029} \\
Moira M. Jardine \orcidlink{0000-0002-1466-5236} \url{https://orcid.org/0000-0002-1466-5236} \\
Luke Bouma \orcidlink{0000-0002-0514-5538} \url{https://orcid.org/0000-0002-0514-5538} \\

\bibliographystyle{mnras}
\bibliography{references}

\label{lastpage}

\end{document}